\pdfoutput=1
\RequirePackage{ifpdf}
\ifpdf % We~are running pdfTeX in pdf mode
\documentclass[pdftex]{sigma}
\else
\documentclass{sigma}
\fi

\numberwithin{equation}{section}

\newcommand{\pslash}{p \! \! \! /}

\newcommand{\xslash}{x \! \! \! /}
\newcommand{\yslash}{y \! \! \! /}
\newcommand{\partialslash}{\partial \! \! \! /}
\newcommand{\half}{\mbox{\small{$\frac{1}{2}$}}}

\newcommand{\threehalves}{\mbox{\small{$\frac{3}{2}$}}}
\newcommand{\MSbar}{\overline{\mbox{MS}}}
\newcommand{\Nc}{N_{\!c}}
\newcommand{\NA}{N_{\!A}}

\begin{document}
\allowdisplaybreaks

\newcommand{\arXivNumber}{2102.12767}

\renewcommand{\thefootnote}{}

\renewcommand{\PaperNumber}{064}

\FirstPageHeading

\ShortArticleName{Generalized Gross--Neveu Universality Class with Non-Abelian Symmetry}

\ArticleName{Generalized Gross--Neveu Universality Class\\ with Non-Abelian Symmetry\footnote{This paper is a~contribution to the Special Issue on Algebraic Structures in Perturbative Quantum Field Theory in honor of Dirk Kreimer for his 60th birthday. The~full collection is available at \href{https://www.emis.de/journals/SIGMA/Kreimer.html}{https://www.emis.de/journals/SIGMA/Kreimer.html}}}

\Author{John A.~GRACEY}

\AuthorNameForHeading{J.A.~Gracey}

\Address{Theoretical Physics Division, Department of Mathematical Sciences, University of Liverpool,\\ P.O.~Box 147, Liverpool, L69 3BX, UK}
\Email{\href{mailto:gracey@liverpool.ac.uk}{gracey@liverpool.ac.uk}}

\ArticleDates{Received February 26, 2021, in final form June 18, 2021; Published online June 29, 2021}

\Abstract{We use the large $N$ critical point formalism to compute $d$-dimensional critical exponents at several orders in $1/N$ in an Ising Gross--Neveu universality class where the core interaction includes a Lie group generator. Specifying a particular symmetry group or taking the abelian limit of the final exponents recovers known results but also provides expressions for any Lie group or fermion representation.}


\Classification{81T17; 81T18; 81V25; 82B27}

\renewcommand{\thefootnote}{\arabic{footnote}}
\setcounter{footnote}{0}

\section{Introduction}

One aspect of quantum field theory that has important applications to Nature is
the study of~fixed points of the renormalization group functions. These are
defined to be the non-trivial zeros of the $\beta$-function. Using the location
of a fixed point one can compute the values of the renormalization group
functions there to produce renormalization scheme independent expressions known
as critical exponents~\cite{1,2,3,4}. These quantities govern the dynamics of~the phase transitions in a material. Indeed accurate measurements of the
exponents experimentally as~well as the symmetry properties of a material, can
equally guide one to the underlying quantum field theory or spin model
describing the dynamics. One property is that more than one theory can be a
valid description of a phase transition. For instance, both a continuum quantum
field theory as well as a discrete spin model with a common symmetry can be
valid tools to provide numerical exponent estimates. The equivalence of both
theoretical techniques at a fixed point is known as universality~\cite{3,4}.
Away from a phase transition each theory will have different properties and be
inequivalent.

One theory that has risen to the fore in this context in recent years is that
of the Ising Gross--Neveu model~\cite{6,5}. This is primarily due to the belief
that it underpins a particular phase transition in graphene. This material is
made up of a sheet of carbon atoms arranged in~a~hexagonal lattice. When the
two dimensional sheet is stretched it can undergo a transition from an
electrical conductor to what is termed a Mott-insulating phase. On the
theoretical side the Ising Gross--Neveu model can be supplemented with
quantum electrodynamics (QED) to describe aspects of other phase transitions.
Equally if the basic or Ising Gross--Neveu model is endowed with extra
symmetries it contributes to the understanding of transitions in other
materials. For example, what is termed the chiral Heisenberg Gross--Neveu model
is an extension of the Gross--Neveu model to include an ${\rm SU}(2)$ symmetry. It is
the effective theory for electrons on a half-filled honeycomb lattice where
there is a phase transition between an anti-ferromagnetic insulating phase and
a semi-metallic one~\cite{8,9,7}. Its criticality properties were studied in~\cite{10}. More recently a variation of this version of the Gross--Neveu model,
called the fractionalized Gross--Neveu model, has been developed~\cite{11}. It
has a novel spectrum that differs from those of other Gross--Neveu models and
has an associated ${\rm SO}(3)$ symmetry. What is clear from this set of emerging
variations of the Gross--Neveu universality class is the common theme of the
core interaction being modified to include a non-abelian symmetry. In this
respect it is completely parallel to the extension of QED to include a
non-abelian symmetry that equates to quantum chromodynamics (QCD) or Yang--Mills
theory with fermions in the fundamental representation of the Lie group
responsible for colour symmetry. The only difference with the Gross--Neveu class
of theories is in the Lorentz structure of the core interaction. As~QCD has
been studied at length using a general Lie group symmetry rather than
specifying ${\rm SU}(3)$ at the outset, which is the group that governs the strong
interactions, it seems sensible to develop a programme to calculate in the
Gross--Neveu model with a parallel non-abelian symmetry. Then the properties of
the various physical applications can be deduced by specifying the appropriate
group parameters.

This is the main task here. We~will consider a generalized Gross--Neveu
universality class with a non-abelian symmetry and calculate the critical
exponents of the theory. This will be achieved by using the critical point
large $N$ formalism pioneered in~\cite{13,12,14} for the nonli\-near~$O(N)$~$\sigma$~model. Here $N$ would correspond to the number of quark flavours in
the analogy with QCD. The elegance of the approach is such that we can deduce
the critical exponents in~$d$ spacetime dimensions as a function of the
non-abelian symmetry group Casimirs. The~fi\-xed point associated with the
formalism is the Wilson--Fisher fixed point in~$d$-dimensions~\cite{3}. As~the
exponents are renormalization group invariants their $\epsilon$ expansion near
$2$ and $4$ dimensions, where~$\epsilon$ measures the difference in these
values from $d$, will agree with the perturbative evaluation of the same
exponents in the respective quantum field theories of the universality class.
More usefully since the dimension of interest for the materials application is
three, one can determine several terms of the $1/N$ series for each exponent.
These provide reasonably accurate estimates for relatively low $N$ when
compared to results from other techniques. However, the benefit of taking
the general non-abelian universality class approach is that estimates will
be readily available if a phase transition with a new symmetry is discovered.

The article is organized as follows. Relevant background concerning the
generalized Gross--Neveu universality class is given in Section~\ref{sec2} together
with the basic critical point large $N$ formalism. Subsequently in Section~\ref{sec3}
we solve the Schwinger--Dyson equations at criticality at~$O\big(1/N^2\big)$ to
produce the fermion anomalous dimension. Various calculational tools that are
necessary for this are reviewed as well. To provide the groundwork for finding
the next order of this exponent, the anomalous dmension exponent of the
bosonic field is determined at~$O\big(1/N^2\big)$ in Section~\ref{sec4}. One of the other
basic exponents in critical systems is that relating to~the correlation
length behaviour and we determine it at $O\big(1/N^2\big)$ in Section~\ref{sec5}. Equipped
with these results, the large $N$ conformal bootstrap formalism at criticality
is applied in Section~\ref{sec6} to~deduce the fermion anomalous dimension at
$O\big(1/N^3\big)$. We~review our results in Section~\ref{sec7} and provide concluding
remarks in Section~\ref{sec8}.

\section{Background}\label{sec2}

To begin with we recall the Lagrangian of the chiral Heisenberg Gross--Neveu--Yukawa theory is~\cite{9}
\begin{gather}
L^{\text{cHGNY}} = {\rm i} \bar{\psi}^{iI} \partialslash \psi^{iI} +
\frac{1}{2} \partial_\mu \tilde{\pi}^a \partial^\mu \tilde{\pi}^a +
g_1 \tilde{\pi}^a \bar{\psi}^{iI} T^a_{IJ} \psi^{iJ} +
\frac{1}{24} g_2^2 \big(\tilde{\pi}^a \tilde{\pi}^a\big)^2 \nonumber
\\ \hphantom{L^{\text{cHGNY}} = }
{} + \frac{1}{24} g_3^2 \tilde{\pi}^a \tilde{\pi}^b
\tilde{\pi}^c \tilde{\pi}^d \mathop{\rm Tr} \big( T^a T^b T^c T^d \big),
\label{lagchgnd4}
\end{gather}
which is renormalizable in four dimensions, where the three couplings $g_1$,
$g_2$ and $g_3$ are dimensionless. This is a generalization of the Lagrangian
studied in~\cite{15} and is in the chiral Heisenberg Gross--Neveu model
universality class. The renormalizability dimension is also termed the cri\-ti\-cal
dimension of the theory. The scalar-fermion interaction includes the group
generator~$T^a$ of~the Lie algebra and the indices take values in the ranges
$1$~$\leq$~$i$~$\leq$~$N$, $1$~$\leq$~$I$~$\leq$~$\Nc$ and
$1$~$\leq$~$a$~$\leq$~$\NA$, where $N$ is the number of flavours of massless
fermions and $\Nc$ and $\NA$ are the respective dimensions of the fundamental
and adjoint representations of the symmetry group. We~note that in~\cite{9} the
specific group considered was ${\rm SU}(2)$. Within our ultimate critical exponents
the generators will manifest themselves through various colour Casimirs such as
$C_F$ and~$C_A$ defi\-ned~by
\begin{gather}
T^a T^a = C_F ,\qquad f^{acd} f^{bcd} = C_A \delta^{ab},
\label{casdef}
\end{gather}
where $f^{abc}$ are the structure constants. The scalar field $\tilde{\pi}^a$
plays a subtle role in the construction of the large $N$ expansion but in four
dimensions it corresponds to a fundamental propagating field. The main aspect
of the large $N$ critical point formalism of~\cite{13,12,14} is that in the
approach to criticality at the Wilson--Fisher fixed point the dynamics are
driven by the core interaction of the universal quantum field theory. For
(\ref{lagchgnd4}) this is the cubic interaction together with the fermion
kinetic term. These two terms determine the canonical dimensions of both fields
by ensuring the action is dimensionless in $d$-dimensions. In effect the
universal Lagrangian at criticality is
\begin{equation*}
L^{\mbox{\footnotesize{cHGN}}} = {\rm i} \bar{\psi} \partialslash \psi +
g \tilde{\pi}^a \bar{\psi} T^a \psi - \frac{1}{2} \tilde{\pi}^a \tilde{\pi}^a ,
%\label{lagchgnprep}
\end{equation*}
where the quadratic term in $\tilde{\pi}^a$ is necessary for large $N$
renormalizability. We~will omit the labels $i$ and $I$ for brevity from now on. We~say in effect since at criticality there is no coupling constant in the
sense it is conventionally used in perturbation theory. So the critical point
universal Lagrangian that will be the foundation for applying the large $N$
critical point formalism of~\cite{13,12} is
\begin{equation}
L^{\mbox{\footnotesize{univ}}} = {\rm i} \bar{\psi} \partialslash \psi +
\pi^a \bar{\psi} T^a \psi - \frac{1}{2g} \pi^a \pi^a,
\label{lagchgnuniv}
\end{equation}
where we have rescaled the scalar field $\tilde{\pi}^a$ to introduce $\pi^a$.
This Lagrangian~(\ref{lagchgnuniv}) is renormalizable in $d$-dimensions in the
large $N$ formalism~\cite{18,16,17,15}, where $1/N$ is the dimensionless
ordering parameter since the perturbative coupling constant is absent at
criticality. Ensuring that the $d$-dimensional Lagrangian~(\ref{lagchgnuniv})
has a dimensionless action means that $\psi$ has canonical dimension
$\half (d-1)$ while that of $\pi^a$ is unity. For~(\ref{lagchgnuniv}) this
implies that $g$ is dimensionless in two dimensions after eliminating the
auxiliary field $\pi^a$ producing
\begin{equation}
L = {\rm i} \bar{\psi}^{iI} \partialslash \psi^{iI} +
\frac{g^2}{2} \big( \bar{\psi}^{iI} T^a_{IJ} \psi^{iJ} \big)^2 .
\label{lagchgnd2}
\end{equation}
This is similar to the Ising Gross--Neveu model discussed in~\cite{19} and to
see the equivalence, one takes the abelian limit of~(\ref{lagchgnd2}) by
replacing the group generators with the unit matrix. This is completely
parallel to taking the abelian limit of QCD to produce QED. The dimensionality
of $\pi^a$ at criticality plays a key role in the connection of the universal
theory and the Lagrangian of~(\ref{lagchgnd4}). In the latter the three
couplings are dimensionless in four dimensions similar to the effective
coupling of the $3$-point interaction of~(\ref{lagchgnuniv}). Therefore
(\ref{lagchgnuniv}) would be strictly non-renormalizable in four dimensions and
the quadratic term in $\pi^a$ would have a dimensionful coupling which would be
the mass. Instead the standard kinetic term and quartic $\pi^a$ interactions of
(\ref{lagchgnd4}) would be relevant. In other words we term the quartic
interactions to be spectator interactions that would be active solely in four
dimensions. Moreover underlying the first two terms of the universal Lagrangian
(\ref{lagchgnuniv}) there are an infinite number of Lorentz scalar operators
built from combinations of $\pi^a$ and its derivatives. A finite subset of
these extra operators would become relevant in even integer dimensions and act
as interim spectators in the infinite tower of renormalizable quantum field
theories that connect to the universal theory in the neighbourhood of their
respective critical dimensions. In this context it is worth noting that the
quartic spectator interactions of~(\ref{lagchgnd4}) play a role in determining
the full fixed point structure of the four dimensional theory. One of these
fixed points though will correspond to the Wilson--Fisher fixed point of the
generalized Gross--Neveu universality class considered here. However, only a
perturbative evaluation of~the four dimensional theory's renormalization group
functions would determine which one it is. While this is beyond the scope of
the present article it is likely to be a solution where the critical values of
all three couplings are non-zero. One non-trivial check in establishing such a~connection though rests in the agreement of the $\epsilon$ expansion of the
various critical exponents with their large $N$ counterparts that will be
determined here.

More concretely we now summarize the key aspects of the large $N$ critical
point formalism for~(\ref{lagchgnuniv}). In the approach to the fixed point the
propagators have the following asymptotic behaviour in coordinate space~\cite{20}
\begin{equation}
\psi(x) \sim \frac{A\xslash}{\big(x^2\big)^\alpha} \big[1 + A^\prime\big(x^2\big)^\lambda\big],\qquad
\pi(x) \sim \frac{C}{\big(x^2\big)^\gamma}\big[1 + C^\prime\big(x^2\big)^\lambda\big],
\label{asympprop}
\end{equation}
where the name of the corresponding field is used. The dimensionless quantities
$A$ and $C$ are the coordinate independent amplitudes that will always occur in
the combination $y=A^2 C$ from the $3$-point interaction. The next to
leading order terms in~(\ref{asympprop}) that involve the exponent~$\lambda$
are called the corrections to scaling. Here $\lambda$ will be identified with
the correlation length exponent~$\nu$ through $1/\nu=2\lambda$. In~addition to the canonical dimension the two fields have anomalous contributions
and the respective full dimensions of $\psi$ and $\pi^a$ are
\begin{equation}
\alpha = \mu + \frac12 \eta ,\qquad
\gamma = 1 - \eta - \chi_\pi,
\label{expdef}
\end{equation}
where we use $d=2\mu$ for shorthand~\cite{12} and $\eta$ and $\chi$ are
the fermion field and vertex anomalous dimensions respectively. For
applications in condensed matter problems the dimension of $\pi^a$ that is
conventionally used is
\begin{equation*}
\eta_\pi = 4 - 2 \mu - 2 ( \eta + \chi_\pi ) .
\end{equation*}
When $\lambda$ corresponds to the correlation length exponent its canonical
dimension will then be taken to be $(\mu-1)$. In this respect the leading and
next to leading order terms of~(\ref{asympprop}) then have different dimensions
which is the reason for the second set of independent dimensionless amplitudes
$A^\prime$ and $C^\prime$. Each of the exponents that we will compute as well
as $y$ will depend on $N$, $\mu$ and the Lie group Casimir invariants. The
dependence on the former means that each entity has a Taylor series in powers
of $1/N$ that is formally given by
\begin{equation}
\eta(\mu) = \sum_{n=1}^\infty \frac{\eta_n(\mu)}{N^n} ,\qquad
y(\mu) = \sum_{n=1}^\infty \frac{y_n(\mu)}{N^n}
\label{expexp}
\end{equation}
for $\eta$ and $y$ for example and we will determine the first three terms of
$\eta$ for~(\ref{lagchgnuniv}).

These general considerations cover the basic formalism for the technique
introduced in~\cite{13,12,14}. To determine all bar $\eta_3$ we can apply the
original method~\cite{13,12} that was used for the Ising Gross--Neveu
universality class in~\cite{22,21,25,26,23,24}. This required solving the
skeleton Schwinger--Dyson equations for the $\psi$ and $\pi^a$ $2$-point
functions. So the scaling forms of these are needed given that
(\ref{asympprop}) represents the critical point behaviour of the propagator.
However one definition of the $2$-point function is that it is the inverse of
the propagator in momentum space and this mapping can be carried out through
the Fourier transform. Using the convention given in~\cite{13,12} which is
\begin{equation}
\frac{1}{\big(x^2\big)^\alpha} = \frac{a(\alpha)}{2^{2\alpha}} \int_k
\frac{{\rm e}^{{\rm i}kx}}{\big(k^2\big)^{\mu-\alpha}},
\label{fourier}
\end{equation}
where $\int_k=\int \frac{{\rm d}^d k}{(2\pi)^d}$ and
\begin{equation}
a(\alpha) \equiv \frac{\Gamma(\mu-\alpha)}{\Gamma(\alpha)} ,
\end{equation}
we transform~(\ref{asympprop}) to momentum space carry out the inversion and
then apply the inverse Fourier transform. This results in the coordinate space
$2$-point function asymptotic scaling forms which are~\cite{20}
\begin{gather}
\psi^{-1}(x) \sim \frac{r(\alpha-1)\xslash}{A\big(x^2\big)^{2\mu-\alpha+1}}
\big[ 1 - A^\prime s(\alpha-1)\big(x^2\big)^\lambda \big], \nonumber
\\
\pi^{-1}(x) \sim \frac{p(\gamma)}{C\big(x^2\big)^{2\mu-\gamma}}
\big[ 1 - C^\prime q(\gamma) \big(x^2\big)^\lambda \big] .
\label{asymp2pt}
\end{gather}
The presence of the function $a(\alpha)$ in the Fourier transform produces a
complicated dependence on $\mu$ and the exponents, leading to the compact
functions
\begin{gather}
p(\gamma) = \frac{a(\gamma-\mu)}{a(\gamma)} ,\qquad
r(\alpha) = \frac{\alpha p(\alpha)}{(\mu-\alpha)}, \nonumber
\\
q(\gamma) = \frac{a(\gamma-\mu+\lambda)a(\gamma-\lambda)}{a(\gamma-\mu)a(\gamma)} ,\qquad
s(\alpha) = \frac{\alpha(\alpha-\mu)q(\alpha)}{(\alpha-\mu+\lambda)(\alpha-\lambda)} .
\end{gather}
While the large $N$ conformal bootstrap formalism of~\cite{14} has its origins
in these asymptotic scaling functions and was applied to the Ising
Gross--Neveu model in~\cite{22,26,23}, the extraction of an expression for
$\eta_3$ derives from the scaling behaviour of the $3$-point function. We~defer
to a later section for the required technicalities of that formalism.

\section{2-point Schwinger--Dyson equation}\label{sec3}

\begin{figure}[ht]\centering
$$
0\ =\ \psi^{-1}\ \ +\ \ \raisebox{-6mm}{\includegraphics[scale=1.5]{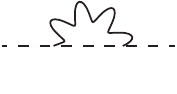}}\ \ +\ \ \raisebox{-6mm}{\includegraphics[scale=1.5]{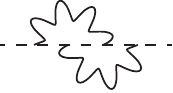}}\,,
\put(-15,-10){\makebox(-0,0)[lb]{$\Sigma_1$}}
$$
$$
0\ =\ \pi^{-1}\ \ +\ \ \raisebox{-6mm}{\includegraphics[scale=1.5]{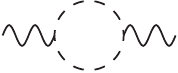}}\ \ +\ \ \raisebox{-6mm}{\includegraphics[scale=1.5]{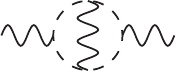}}\,.
\put(-15,-17){\makebox(-0,0)[lb]{$\Pi_1$}}
$$
\caption{$O\big(1/N^2\big)$ corrections to the skeleton Schwinger--Dyson $2$-point functions.}\label{figsde2pt}
\end{figure}

Equipped with the asymptotic scaling forms of the full propagators,
(\ref{asympprop}) and~(\ref{asymp2pt}), which represent the behaviour at
criticality, we use them to solve the Schwinger--Dyson equations. In~conventional perturbation theory one systematically renormalizes the divergent
$n$-point Green's functions in a renormalizable theory order by order in
perturbation theory. This principle is respected in the large $N$ technique of
\cite{13,12} except that the ordering of graphs in the $n$-point functions
is achieved by the variable $1/N$ which is dimensionless across all spacetime
dimensions unlike the perturbative coupling constant. For~(\ref{lagchgnuniv})
the first few terms in the respective $2$-point functions of the fields are
given in Figure~\ref{figsde2pt}, where the dotted line represents the fermion
and the wiggly line denotes the $\pi^a$ field. The two loop graphs are the
$O\big(1/N^2\big)$ corrections to the one loop ones. The counting of powers of $N$
arise from closed fermion loops giving a factor of $N$ and the $\pi^a$ field.
The expansion of the amplitude variable begins at $O(1/N)$ and this translates
into each $\pi^a$ line in Figure~\ref{figsde2pt} carrying a power of $1/N$.
One key point worth noting concerns the lack of dressing of lines with
self-energy corrections. Contributions from such graphs are already accounted
for in the inclusion of a non-zero anomalous dimension in the power of the
asymptotic scaling forms.

At leading order the two equations of Figure~\ref{figsde2pt} equate to
\begin{gather}
0 = r(\alpha-1) + C_F y + O \bigg(\frac{1}{N^2} \bigg), \nonumber
\\
0 = p(\gamma) + T_F N y + O \bigg(\frac{1}{N} \bigg),
\label{sde2pt1}
\end{gather}
where we have included the respective group theory factors which derive from
the properties of~$T^a$ and
\begin{equation*}
\mathop{\rm Tr} \big( T^a T^b \big) = T_F \delta^{ab} .
\end{equation*}
In this coordinate space representation the one loop graphs require no
evaluation. This is because one integrates over the coordinate of the internal
vertices. As~the one loop graphs have external vertices the corresponding terms
of~(\ref{sde2pt1}) are the products of the propagators. In this leading order
instance any integration has been effected in the derivation of the scaling
forms for the {\em full} $2$-point functions. The algebraic representation of
Figure~\ref{figsde2pt} is~(\ref{sde2pt1}) and to $O(1/N)$ it contains two
unknowns. Using~(\ref{expexp}) and the $1/N$ expansion of $r(\alpha-1)$ and
$p(\gamma)$ these are $\eta_1$ and $y_1$ and moreover at this order they occur
linearly. Thus eliminating $y_1$ between the equations of~(\ref{sde2pt1})
produces
\begin{equation*}
\eta_1 = - \frac{2 \Gamma(2\mu-1) C_F}{\mu\Gamma(1-\mu)\Gamma(\mu-1)
\Gamma^2(\mu)T_F} .
\end{equation*}

At next order the situation is not as straightforward due to the additional
graphs of Figure~\ref{figsde2pt} being divergent which necessitates the
introduction of renormalization constants. The formalism for this was provided
in~\cite{16,17} and requires the introduction of a regularization which is
achieved through the shift~\cite{13,12}
\begin{equation}
\chi_\pi \rightarrow \chi_\pi + \Delta,\label{chiDel}
\end{equation}
where $\Delta$ is a small parameter. In effect it equates to an analytic
regularization of the propagators and we emphasize that the spacetime dimension
$d$ does not play any role in the regularization in contrast to coupling
constant perturbation theory. Consequently the extension of~(\ref{sde2pt1}) to
the next order has to account for this and so the algebraic representation of
Figure~\ref{figsde2pt} becomes
\begin{gather}
0 = r(\alpha-1) + C_F y Z_V^2 \big(x^2\big)^{\chi_\pi+\Delta} +
\frac12 C_F [ 2 C_F - C_A ] y^2 Z_V^4 \Sigma_1 \big(x^2\big)^{2\chi_\pi+2\Delta} +
O \bigg( \frac{1}{N^3} \bigg), \nonumber\\
0 = p(\gamma) + T_F N y Z_V^2 \big(x^2\big)^{\chi_\pi+\Delta} +
\frac12 T_F [ 2 C_F - C_A ]N y^2 Z_V^4 \Pi_1 \big(x^2\big)^{2\chi_\pi+2\Delta} +
O \bigg( \frac{1}{N^2} \bigg)\label{sde2pt2}
\end{gather}
in coordinate space where $Z_V$ is the vertex renormalization constant. It has
the Laurent expansion
\begin{equation*}
Z_V = 1 + \sum_{l=1}^\infty \sum_{n=1}^l \frac{m_{ln}}{\Delta^n},
\end{equation*}
where the residues are
\begin{equation*}
m_{ln} = \sum_{i=1}^\infty \frac{m_{ln,i}}{N^i},
\end{equation*}
after expanding in powers of $1/N$. We~follow~\cite{16,17} and restrict to the
$\MSbar$ scheme. In~(\ref{sde2pt2}) the~$x^2$ dependence arises from the
dimensionality of the integrals in the regularized Lagrangian. As~it stands
the various factors prevent the limit to criticality from being taken
smoothly. Moreover the factors associated with the one loop graphs of Figure
\ref{figsde2pt} will give contributions at~$O\big(1/N^2\big)$ from the expansion of
$\big(x^2\big)^{\chi_\pi}$. This will produce problematic logarithms but these are
connected with the simple poles of the values of the two loop graphs denoted by
$\Sigma_1$ and $\Pi_1$. In~particular they have the formal structure
\begin{equation*}
\Sigma_1 = \frac{K_1}{\Delta} + \Sigma_1^\prime ,\qquad
\Pi_1 = \frac{L_1}{\Delta} + \Pi_1^\prime,
\end{equation*}
where $\Sigma^\prime_1$ and $\Pi^\prime_1$ are finite. They were computed
previously in~\cite{20}, where the explicit $d$-de\-pendent values are available. We~note our trace conventions at this stage are the same as~\cite{20} and we
use $2$~$\times$~$2$ $\gamma$-matrices. To adjust for higher dimensional
$\gamma$-matrix representations one simply redefines $N$ using
\begin{equation*}
N = \frac{1}{2} d_\gamma N,
\end{equation*}
where $d_\gamma$ is the dimension of the $\gamma$-matrix representation.
\begin{figure}[ht]\centering
\[\qquad
\raisebox{-10mm}{\includegraphics[scale=1.5]{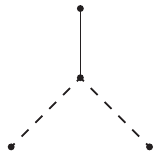}}\hspace{60mm}
\raisebox{-10mm}{\includegraphics[scale=1.5]{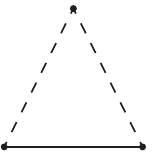}}
\put(-233,-9){\makebox(-0,0)[lb]{$\equiv\ \dfrac{a(\alpha)a(\beta-1)a(\gamma-1)}{(\beta-1)(\gamma-1)}
\quad\ \mu-\beta+1\qquad\qquad \mu-\gamma+1$}}
\put(-273,40){\makebox(-0,0)[lb]{$0$}}
\put(-36,40){\makebox(-0,0)[lb]{$0$}}
\put(-305,-35){\makebox(-0,0)[lb]{$x$}}
\put(-68,-35){\makebox(-0,0)[lb]{$x$}}
\put(-243,-37){\makebox(-0,0)[lb]{$y$}}
\put(-7,-37){\makebox(-0,0)[lb]{$y$}}
\put(-47,-40){\makebox(-0,0)[lb]{$\mu-\alpha$}}
\put(-273,-9){\makebox(-0,0)[lb]{$z$}}
\put(-268,17){\makebox(-0,0)[lb]{$\alpha$}}
\put(-293,-10){\makebox(-0,0)[lb]{$\gamma$}}
\put(-251,-10){\makebox(-0,0)[lb]{$\beta$}}
\]
\caption{Uniqueness rule for scalar-fermion vertex for arbitrary exponents
$\alpha$, $\beta$ and $\gamma$.}\label{figuniq}
\end{figure}

The key integration tool for evaluating the graphs in~\cite{20} is shown in
Figure~\ref{figuniq} and is termed uniqueness or conformal integration for the
scalar-Yukawa interaction. There are several ways to establish the relation
provided in Figure~\ref{figuniq}. One is to use Feynman parameters. In that
derivation the final integration is over a hypergeometric function and it
cannot proceed unless the uniqueness condition of
\begin{equation*}
\alpha + \beta + \gamma = 2 \mu + 1
\end{equation*}
is fulfilled. Setting this allows the final integration to be completed which
produces the factor on the right side of Figure~\ref{figuniq}. A more elegant
alternative is to apply a conformal transformation to the integral which is
\begin{equation}
x_\mu \to \frac{x_\mu}{x^2},\qquad
y_\mu \to \frac{y_\mu}{y^2},\qquad
z_\mu \to \frac{z_\mu}{z^2},
\label{confmap1}
\end{equation}
which implies the mapping
\begin{equation}
( \xslash - \yslash ) \to - \frac{\yslash ( \xslash - \yslash ) \xslash}{x^2 y^2}
 = - \frac{\xslash ( \xslash - \yslash ) \yslash}{x^2 y^2}
\label{confmap2}
\end{equation}
for instance. The consequence is that when applied to strings of contracted
$\gamma$-matrices the transformation does not alter the initial string of
$\gamma$-matrices. In the application to the left hand side of the equation of
Figure~\ref{figuniq} the exponent of the scalar becomes
$(2\mu+1-\alpha-\beta-\gamma)$. Setting this to zero allows the $z$-integration
to proceed resulting in the expression on the right hand side after undoing the
initial conformal transformation.

With the availability of the counterterm from the vertex renormalization
constant the divergences are removed minimally. However $\ln\big(x^2\big)$ terms remain
through two contributions in each Schwinger--Dyson equation. One is from the
power of $\big(x^2\big)^{2\Delta}$ in the $O\big(1/N^2\big)$ correction. The other arises from
the factor $\big(x^2\big)^{\chi_\pi}$ that is present in the one loop graph. Expanding
this in powers of $1/N$ the $O(1/N)$ term involves $\ln\big(x^2\big)$. To be able to
take the $x^2$~$\to$~$0$ limit safely means that the as yet undefined
$\chi_{\pi\,1}$ has to be suitably chosen. Doing so to ensure there are no
$\ln\big(x^2\big)$ terms in each Schwinger--Dyson equation at $O\big(1/N^2\big)$ requires
\begin{equation}
\chi_{\pi\, 1} = \frac{(2C_F - C_A) \mu}{2 (\mu-1) C_F} \eta_1
\label{chi1}
\end{equation}
from the explicit values for the residues which satisfy $L_1=-$~$2K_1$
and implies the same value results for both equations. This finally renders
the algebraic representation of Figure~\ref{figsde2pt} finite as well as
ensuring that it is scale free. Since the two equations have two unknown
variables $\eta_2$ and $y_2$ that appear linearly, eliminating the latter leads
to
\begin{equation*}
\eta_2 = \bigg[ \frac{(2 \mu-1) C_F}{(\mu-1)} \Psi(\mu)
- \frac{\mu C_A}{2 (\mu-1)} \Psi(\mu)
+ \frac{(4 \mu-1) (2 \mu-1) C_F}{2 \mu (\mu-1)^2}
- \frac{3 \mu C_A}{4 (\mu-1)^2} \bigg] \frac{\eta_1^2}{C_F},
\end{equation*}
where we use the shorthand
\begin{equation*}
\Psi(\mu) = \psi(2\mu-1) - \psi(1) + \psi(2-\mu) - \psi(\mu)
\end{equation*}
which involves the Euler $\psi$ function defined by
$\psi(z)= {\rm d} \ln \Gamma(z)/{\rm d}z$.

\section[pi a critical exponent at O(1/N2)]
{$\boldsymbol{\pi^a}$ critical exponent at $\boldsymbol{O\big(1/N^2\big)}$}\label{sec4}

\begin{figure}[ht]\centering
\includegraphics[scale=1.1]{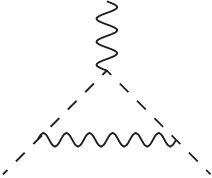}

\caption{$O(1/N)$ corrections to vertex function.}\label{figchi1}
\end{figure}

\begin{figure}[ht]\centering
$$
\raisebox{-10mm}{\includegraphics{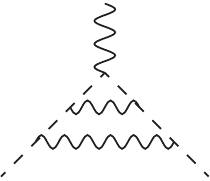}}\quad +\quad
\raisebox{-10mm}{\includegraphics{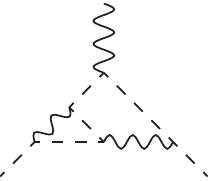}}\quad +\quad
\raisebox{-10mm}{\includegraphics{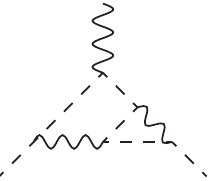}}\quad +\quad
\raisebox{-10mm}{\includegraphics{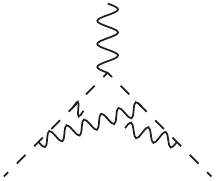}}
$$
$$
+\quad\raisebox{-10mm}{\includegraphics{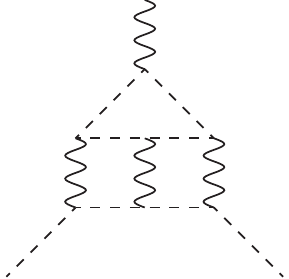}}\quad +\quad
\raisebox{-10mm}{\includegraphics{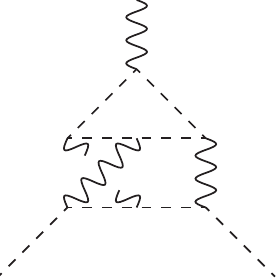}}\quad +\quad
\raisebox{-10mm}{\includegraphics{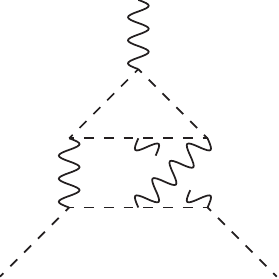}}
$$
\caption{$O\big(1/N^2\big)$ corrections to vertex function.}\label{figchi2}
\end{figure}

Having established the fermion critical exponent at $O\big(1/N^2\big)$ the next stage
in the large $N$ formalism is to determine the same quantity for the boson
field. In this instance from the definition~(\ref{expdef}) this requires the
vertex anomalous dimension at $O\big(1/N^2\big)$. While $\chi_{\pi\,1}$ followed as a
corollary to ensuring the $2$-point function was finite in the approach to
criticality, in order to proceed to the next order to find $\chi_{\pi\,2}$ by
the same method is too intractable. Indeed evaluating the analogous exponent in
other models has not been achieved that way. Instead a more direct approach
suffices which is to examine the scaling behaviour of the $3$-point vertex in
the critical limit. In other words the $O\big(1/N^2\big)$ graphs illustrated in Figures
\ref{figchi1} and~\ref{figchi2} are computed.

In practical terms the diagrams are more straightforward to evaluate in
momentum space than in coordinate space. This is primarily due to
simplifications in the application of the uniqueness rule of~\cite{20}. However
one can connect the values of graphs in both the coordinate and momentum space
evaluations through the Fourier transform~(\ref{fourier}). Underlying this one
needs to use the momentum space forms of the asymptotic scaling functions which
are
\begin{equation}
\psi(p) \sim \frac{\tilde{A}\pslash}{(p^2)^{\mu-\alpha+1}} ,\qquad
\pi(p) \sim \frac{\tilde{C}}{(p^2)^{\mu-\gamma}},\label{asymppropmom}
\end{equation}
where we have new momentum independent amplitudes $\tilde{A}$ and $\tilde{C}$
with the associated variable $\tilde{y}=\tilde{A}^2 \tilde{C}$. The
explicit value for the latter can be related to the known expansion of $y$
either via the Fourier transform relation of~(\ref{asympprop}) and
(\ref{asymppropmom}) or by repeating the exercise of the previous section by
setting up the formalism in momentum space at the outset. Both approaches lead
to~the same expression for $\tilde{y}$ to $O\big(1/N^2\big)$ as well as $\eta_2$ as a
check. In relation to this momentum space $2$-point function approach the
asymptotic scaling forms of the $2$-point functions are
\begin{equation*}
\psi^{-1}(p) \sim \frac{\pslash}{\tilde{A}(p^2)^{\alpha-\mu}} ,\qquad
\pi^{-1}(p) \sim \frac{1}{\tilde{C} (p^2)^{\gamma-\mu}} .
\end{equation*}
It is the inverse Fourier transform of these that produce the leading terms of~(\ref{asymp2pt}).

With~(\ref{asymppropmom}) it is straightforward to evaluate the graph of Figure~\ref{figchi1} and determine an expression for $\chi_{\pi\,1}$ from the scaling
behaviour. It is equivalent to~(\ref{chi1}). The key point is that the same
procedure of~\cite{16,17} can now be applied to determine $\chi_{\pi\,2}$. This
requires in part the evaluation of the $O\big(1/N^2\big)$ corrections illustrated in
Figure~\ref{figchi2}, where the fermions can be routed around the closed loop in
both directions. The values of the diagrams in the absence of any group theory
considerations were given in~\cite{21}. With the presence of the group
generator in the vertex of~(\ref{lagchgnuniv}), obtaining the associated group
factor of each graph is straightforward in most cases using~(\ref{casdef}) for
example and the definition of the Lie algebra which in our conventions is
\begin{equation*}
\big[ T^a, T^b \big] = {\rm i} f^{abc} T^c .
\end{equation*}
However for the graphs where there is a closed fermion loop with four $\pi^a$
fields attached a higher group Casimir is present. In particular the group
factor associated with each graph will contain the fully symmetric rank $4$
tensor
\begin{equation*}
d_F^{abcd} = \frac{1}{6} \mathop{\rm Tr} \big( T^a T^{(b} T^c T^{d)} \big)
\end{equation*}
among others. To treat these so called light-by-light graphs we have used the
{\tt color.h} routine that accompanies the symbolic manipulation language
{\sc Form}~\cite{27,28}. The package allows one to manipulate group theory
factors associated with Feynman graphs and write them in terms of~Casimirs. It~is based on the comprehensive analysis provided in~\cite{29}. However, rather
than use {\tt color.h} to determine the group factor solely for these
light-by-light graphs we have applied it to all the graphs treated throughout
this article for consistency. We~note that the rank $3$ fully symmetric colour
tensor $d^{abc}$ given by\vspace{-1ex}
\begin{gather*}
d^{abc} = \frac{1}{2} \mathop{\rm Tr} \big( T^a T^{(b} T^{c)} \big)
\end{gather*}
also occurs in graphs that contain $d_F^{abcd}$. In the final expressions for
the exponents, however, these are absent as a result of a cancellation between
graphs where the fermions are routed around the closed loop in different
directions. In addition to the $O\big(1/N^2\big)$ corrections of Figure~\ref{figchi2}
there are contributions to $\chi_{\pi\,2}$ from the graph of Figure~\ref{figchi1}. These arise from the correction to the variable $\tilde{y}$
which is $\tilde{y}_2$ as well as the parts from the $1/N$ expansion of the
exponents of the propagators in~(\ref{asymppropmom}). In addition one has to
include the vertex renormalization constant $Z_V$ in~the $O(1/N)$ graph as this
cancels the subgraph divergences in the first three graphs of the first row in
Figure~\ref{figchi2}. This cancellation is necessary in order to ensure the
approach to criticality is smooth. Assemblying all these components allows us
to deduce the vertex critical exponent at~the next order which is
\begin{gather}
\chi_{\pi\,2} = \bigg[ \frac{\mu (2 \mu-1)}{(\mu-1)^3} C_F^2
- \frac{\mu (7 \mu-2)}{4(\mu-1)^3} C_F C_A
- \frac{\mu^2 (2 \mu^2-6 \mu+1)}{12(\mu-1)^3} C_A^2
\nonumber
\\ \hphantom{\chi_{\pi\,2} =\bigg[}
{}- \frac{\mu^2 (2 \mu-1)}{(\mu-1)^2} \frac{d_F^{abcd} d_F^{abcd}}{N_A T_F^2}
+ \frac{\mu ( C_A - 2 C_F ) ( C_A \mu - 2 (2 \mu-1) C_F )}{4(\mu-1)^2} \Psi(\mu)
\nonumber
\\ \hphantom{\chi_{\pi\,2} =\bigg[}
{}- \frac{ \mu^2 ( C_A^2 T_F^2 N_A - 24 d_F^{abcd} d_F^{abcd} )}{8(\mu-1)
N_A T_F^2} \Theta(\mu)\bigg] \frac{\eta_1^2}{C_F^2} ,
\label{chipi2}
\end{gather}
where\vspace{-1ex}
\begin{gather*}
\Theta(\mu) = \psi^\prime(\mu) - \psi^\prime(1),
\end{gather*}
and the contributions from the light-by-light graphs is evident. We~close this
section by noting that the $d_F^{abcd}$ tensor will appear in the same
combination as it does in~(\ref{chipi2}) in the perturbative renormalization
group functions of~(\ref{lagchgnd4}) from Feynman diagrams involving
interactions with the couplings $g_1$ and $g_3$.\vspace{-1ex}

\section[Correlation length exponent at O(1/N2)]
{Correlation length exponent at $\boldsymbol{O\big(1/N^2\big)}$}\label{sec5}\vspace{-1ex}

Having established the dimensions of the two fields at $O\big(1/N^2\big)$ using the
leading term of the asymptotic forms of the propagators in the approach to
criticality, it is possible to study the corrections to scaling. These are
contained in both~(\ref{asympprop}) and~(\ref{asymp2pt}) corresponding to the
terms involving the coordinate independent additional dimensionless parameters
$A^\prime$ and $C^\prime$. The extra exponent $\lambda$ can be regarded as any
exponent but to access the correlation length exponent $\nu$ we set
$\lambda=1/(2\nu)$ which has the canonical dimension of $(\mu-1)$. To
determine the $1/N$ corrections in $d$-dimensions to this exponent requires a
consistency equation that extends~(\ref{sde2pt1}) and~(\ref{sde2pt2}). To find
$\lambda_1$ we substitute the various asymptotic scaling functions into the
Schwinger--Dyson equations for the $2$-point function which produces the
representation
\begin{gather}
0 = r(\alpha-1) \big[ 1 - A^\prime s(\alpha-1)\big(x^2\big)^\lambda \big] +
C_F y Z_V^2 \big(x^2\big)^{\chi_\pi+\Delta}\big[1 + ( A^\prime + C^\prime) \big(x^2\big)^\lambda\big]
\nonumber
\\ \hphantom{0 =}
{} + \frac12 C_F [ 2 C_F - C_A ] y^2 \big(x^2\big)^{2\chi_\pi+2\Delta}
\big[ \Sigma_1 + \big( \Sigma_{1A} A^\prime + \Sigma_{1C} C^\prime
\big) \big(x^2\big)^\lambda \big] + O \bigg( \frac{1}{N^3} \bigg)
\label{psisdelam}
\end{gather}
and
\begin{gather}
0 = p(\gamma) \big[ 1 - C^\prime q(\gamma)\big(x^2\big)^\lambda \big] +
T_F N y Z_V^2 \big(x^2\big)^{\chi_\pi+\Delta} \big[ 1 + 2 A^\prime \big(x^2\big)^\lambda\big] \nonumber
\\ \hphantom{0 =}
{} - \frac12 T_F [ 2 C_F - C_A ] N y^2 \big(x^2\big)^{2\chi_\pi+2\Delta}
\big[ \Pi_1 + \big( \Pi_{1A} A^\prime + \Pi_{1C} C^\prime\big)
\big(x^2\big)^\lambda \big] + O \bigg( \frac{1}{N^2} \bigg),
\label{pisdelam}
\end{gather}
where we have omitted the factor of $Z_V^4$ in the respective $O\big(1/N^2\big)$ and
$O(1/N)$ correction terms. The counterterms from these only come into effect at
the next order.

Unlike the computation of $\eta_1$ we have included the two loop graphs of
Figure~\ref{figsde2pt}, where the correction to scaling is included. These are
denoted by $\Sigma^\prime_{1\phi}$ and $\Pi^\prime_{1\phi}$, where $\phi$
indicates that the insertion is on either a $\psi$ or $\pi^a$ line according to
whether it is~$A$ or~$C$ respectively. The reason why these all have to be
included in principle resides in the leading order~$N$ dependence of the
$2$-point scaling functions. For the two key combinations that appear in the
correction to scaling Schwinger--Dyson equation we note that this dependence is~\cite{22,24}
\begin{equation}
r(\alpha-1) s(\alpha-1) = O(1) ,\qquad
p(\gamma) q(\gamma) = O \bigg( \frac{1}{N} \bigg) .
\label{lofuns}
\end{equation}
In fact the constant of proportionality of the latter is the combination
$(\lambda_1-\eta_1-\chi_{\pi\,1})$. As~$\eta_1$ and $\chi_{\pi\,1}$ are both
available this leaves $\lambda_1$ as the unknown we seek. These terms will
form part of the consistency equation that determines $\lambda$ to $O\big(1/N^2\big)$
and emerges from decoupling the $\big(x^2\big)^\lambda$ terms in~(\ref{psisdelam}) and~(\ref{pisdelam}) which follows on dimensional grounds. The resulting two
equations are
\begin{gather*}
0 = - r(\alpha-1) s(\alpha-1) A^\prime +
C_F y Z_V^2 \big(x^2\big)^{\chi_\pi+\Delta} \big[ A^\prime + C^\prime \big]\nonumber
\\ \hphantom{0 =}
{} + \frac12 [ 2 C_F - C_A ] y^2 \big(x^2\big)^{2\chi_\pi+2\Delta}
\big[ \Sigma_{1A} A^\prime+ \Sigma_{1C} C^\prime \big] + O \bigg( \frac{1}{N^3} \bigg)
%\label{psilam}
\end{gather*}
and
\begin{gather*}
0 = - p(\gamma) q(\gamma) C^\prime +
T_F N y Z_V^2 \big(x^2\big)^{\chi_\pi+\Delta} A^\prime \nonumber
\\ \hphantom{0 =}
{} - \frac12 C_F [ 2 C_F - C_A ] N y^2 \big(x^2\big)^{2\chi_\pi+2\Delta}
\big[ \Pi_{1A} A^\prime + \Pi_{1C} C^\prime \big] +
O \bigg( \frac{1}{N^2} \bigg).
%\label{pilam}
\end{gather*}
Alternatively the relevant terms that produce an expression for $\lambda_1$
with respect to the large $N$ expansion due to~(\ref{lofuns}) can be written
as a matrix ${\cal M}$, where
\begin{equation*}
{\cal M} =
\begin{pmatrix}
- r(\alpha-1) s(\alpha-1) C_F y \\[1ex]
T_F N y - p(\gamma) q(\gamma) -
\frac12 C_F [ 2 C_F - C_A ] N y^2 \Pi^\prime_{1C}
\end{pmatrix}\!.
%\label{mat1}
\end{equation*}
Examining the $(2,2)$ element both terms are the same order in $1/N$ as $y$ is
$O(1/N)$. Setting $\det {\cal M}=0$ produces the consistency equation
for $\lambda_1$ which can be solved to give
\begin{equation}
\lambda_1 = - (2\mu-1) \eta_1 .\label{explam1}
\end{equation}

\begin{figure}[ht]\centering
$$
\hphantom{+\quad}\raisebox{-8mm}{\includegraphics[scale=1.05]{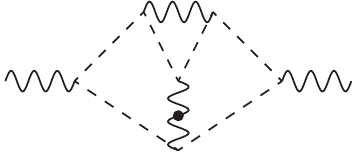}}\quad +\quad
\raisebox{-8mm}{\includegraphics[scale=1.05]{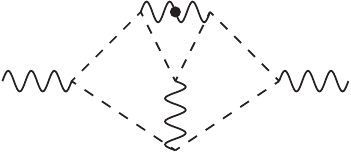}}
$$
$$
+\quad\raisebox{-8mm}{\includegraphics[scale=1.05]{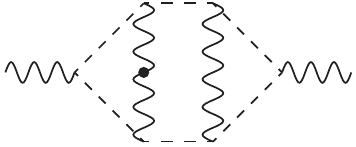}}\quad+\quad
\raisebox{-8mm}{\includegraphics[scale=1.05]{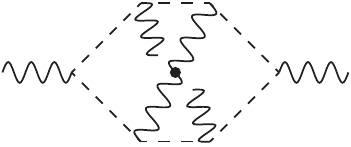}}
$$
$$
+\quad\raisebox{-8mm}{\includegraphics[scale=1.05]{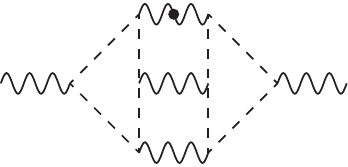}}\quad+\quad
\raisebox{-8mm}{\includegraphics[scale=1.05]{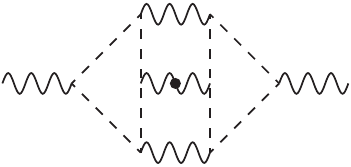}}
$$
$$
+\quad\raisebox{-8mm}{\includegraphics[scale=1.05]{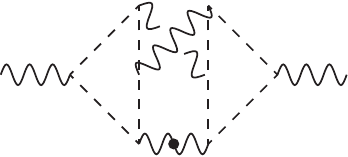}}\quad +\quad
\raisebox{-8mm}{\includegraphics[scale=1.05]{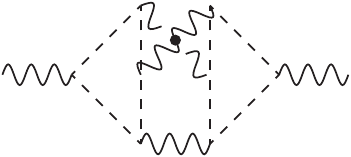}}
$$
\caption{Higher order graphs for corrections to the $\pi^a$ Schwinger--Dyson $2$-point function needed for~$\lambda_2$.}
\label{figlam2}
\end{figure}

To proceed to the next stage and find $\lambda_2$ the higher order $1/N$
correction graphs have to be added in to the two Schwinger--Dyson equations that
produced the $O(1/N)$ consistency equations. However the ones we neglected to
determine~(\ref{explam1}), since their $N$ dependence in the determinant is an
order lower that the contribution from $\Pi_{1C}^\prime$, now have to be
included. These are $\Sigma_{1A}^\prime$, $\Sigma_{1C}^\prime$ and
$\Pi_{1A}^\prime$ while the graph corresponding to $\Pi_{1C}^\prime$ has to be
expanded to the next order in $1/N$ since there is $N$ dependence in the
propagator exponents. To ease the extraction of the expansion of the
consistency equation determinant we formally set
\begin{equation*}
\Pi^\prime_{1C} = \Pi^\prime_{1C1} + \Pi^\prime_{1C2} \frac{1}{N} +
O \bigg( \frac{1}{N^2} \bigg)
\end{equation*}
to clarify this. The main work however resides in including the final
contributions to find $\lambda_2$ which are illustrated in Figure~\ref{figlam2}. While the individual $d$-dependent values have been recorded in~\cite{25} for instance, we have had to append the respective group theory
factors. Again we have used the {\sc Form} {\tt color.h} routine due to the
presence of the light-by-light diagrams. Repeating the exercise of setting the
determinant of the consistency equations to zero at the next order in~$1/N$
produces the expression
\begin{gather*}
\lambda_2 = \bigg[\bigg[ \frac{\mu \big(3 \mu^2-6 \mu+2\big) C_A^2 C_F}{6 T_F}
+ 4 \mu \frac{d_F^{abcd} d_F^{abcd} C_F}{T_F^3N_A} \bigg]\frac{1}{(\mu-1) (\mu-2)^2 \eta_1}
\nonumber
\\ \hphantom{\lambda_2 = }
{}- \bigg[ \frac{1}{24} \mu^2 (2 \mu-3) C_A^2
+ 2 \mu^2 (2 \mu-3) \frac{d_F^{abcd} d_F^{abcd}}{T_F^2 N_A} \bigg]
\frac{[\Psi^2(\mu) + \Phi(\mu)]}{(\mu-1)(\mu-2)}
\nonumber
\\ \hphantom{\lambda_2 = }
{}+ \bigg[{-}(2 \mu-1)^2 (\mu+1) (\mu-1) (\mu-2)^2 C_F^2
+ \mu (2 \mu-1) (\mu-1) (\mu-2)^2 C_F C_A
\nonumber
\\ \hphantom{\lambda_2 = \bigg[}
{}+ \frac{1}{24} \mu^2 (\mu\!-1) \big(6 \mu^2\!-21 \mu+20\big) C_A^2
\!- \mu^2 (3 \mu\!-5) (2 \mu\!-5) \frac{d_F^{abcd} d_F^{abcd}}{T_F^2 N_A} \bigg]
\frac{\Psi(\mu)}{(\mu\!-1)^2 (\mu\!-2)^2}
\nonumber
\\ \hphantom{\lambda_2 = }
{}+ \bigg[{-} \frac{3}{2} \mu^2 (2 \mu+1) (\mu-2) C_F^2
+ \frac{3}{4} \mu^2 (2 \mu+5) (\mu-2) C_F C_A- \frac{11}{8} \mu^2 (\mu-2) C_A^2
\nonumber
\\ \hphantom{\lambda_2 = \bigg[}
{}+ 3 \mu^2 (5 \mu-7) \frac{d_F^{abcd} d_F^{abcd}}{T_F^2 N_A}\bigg]
\frac{\Theta(\mu)}{(\mu-1) (\mu-2)}+ \frac{3 \mu (2 \mu-1)}{4 (\mu-1)^2} C_F C_A
\nonumber
\\ \hphantom{\lambda_2 = }
{}+ \frac{(2 \mu-1)^2 \big(2 \mu^3-4 \mu^2-2 \mu+1\big)}{2 \mu (\mu-1)^2} C_F^2
- \frac{\mu^2 \big(8 \mu^4-42 \mu^3+85 \mu^2-75 \mu+20\big)}{48 (\mu-1)^3 (\mu-2)^2} C_A^2
\nonumber
\\ \hphantom{\lambda_2 = }
{}- \frac{\mu^2 \big(4 \mu^4-18 \mu^3+26 \mu^2-15 \mu+7\big)}
{2 (\mu-1)^3 (\mu-2)^2} \frac{d_F^{abcd} d_F^{abcd}}{T_F^2 N_A}\bigg] \frac{\eta_1^2}{C_F^2} .
\end{gather*}
We have introduced the additional shorthand notation
\begin{equation*}
\Phi(\mu) = \psi^\prime(2\mu-1) - \psi^\prime(2-\mu) -
\psi^\prime(\mu) + \psi^\prime(1) .
\end{equation*}
Essential in determining this was the values of $\eta_2$ and $\chi_{\pi\,2}$.

\section[Large N conformal bootstrap]
{Large $\boldsymbol{N}$ conformal bootstrap}\label{sec6}

\begin{figure}[ht]\centering
$$
\raisebox{-8mm}{\includegraphics{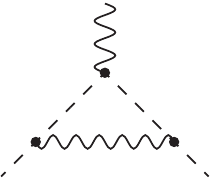}}\quad +\quad
\raisebox{-8mm}{\includegraphics{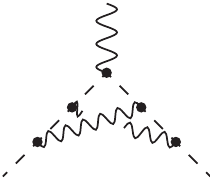}}
$$
$$
+\quad\raisebox{-10mm}{\includegraphics{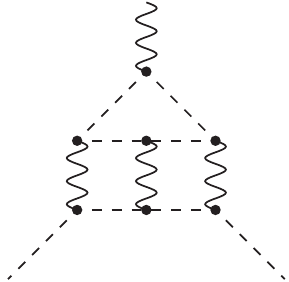}}\quad +\quad
\raisebox{-10mm}{\includegraphics{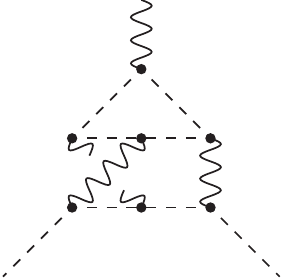}}\quad +\quad
\raisebox{-10mm}{\includegraphics{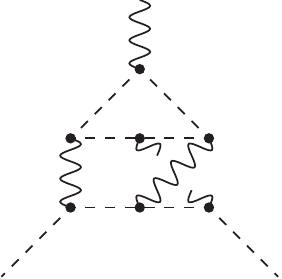}}
$$
\caption{Primitive graphs which determine $\eta_3$ in the large $N$ conformal bootstrap method.}
\label{figeta3sde}
\end{figure}

\begin{figure}[ht]\centering
\includegraphics{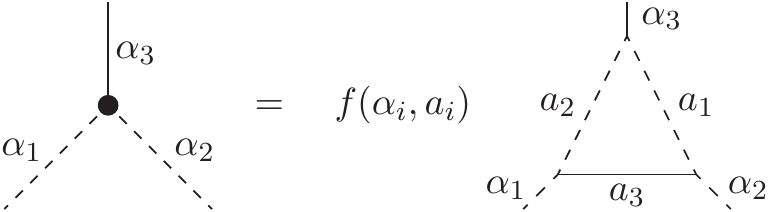}

\caption{Polyakov conformal triangle for a scalar Yukawa interaction.}\label{figcntr}
\end{figure}

The final part of our study follows a different tack by applying the large $N$
conformal bootstrap programme developed in~\cite{14}, based on the early
insights of~\cite{33,34,32,30,31}. In general terms the focus is initially
directed towards the $3$-point vertex of~(\ref{lagchgnuniv}) and its behaviour
in the critical region. The fields will still obey the asymptotic scaling forms
of~(\ref{asymp2pt}) but in treating the Green's functions in the bootstrap
formalism not only are there no self-energy corrections on the propagators but
there are no vertex corrections. In effect the primitive graphs are the
building blocks and are illustrated in Figure~\ref{figeta3sde}. In that figure
the dotted vertices do not correspond to the vertex of the Lagrangian
(\ref{lagchgnuniv}). Instead they denote the presence of a Polyakov conformal
triangle~\cite{30} which includes all vertex corrections at criticality. It is
defined in Figure~\ref{figcntr} for the general Yukawa type interaction that
includes the one of~(\ref{lagchgnuniv}). The external exponents $\alpha_i$ are
general and are determined from the underlying theory. For example
$\alpha_1=\alpha_2=\alpha$ and $\alpha_3=\gamma$ for
(\ref{lagchgnuniv}). The values of the internal indices $a_i$ are the solution
to the simultaneous equations
\begin{gather*}
a_1 + a_2 + \alpha_3 = 2\mu + 1, \nonumber \\
a_2 + a_3 + \alpha_1 = 2\mu + 1, \nonumber \\
a_3 + a_1 + \alpha_2 = 2\mu + 1 .
\end{gather*}
They ensure that the internal vertices of the triangle graph in Figure
\ref{figcntr} are all unique unlike the vertex on the left side of the equation
for~(\ref{lagchgnuniv}). The calculational benefit of regarding the full vertex
correction as a conformal triangle is that applying a conformal transformation,
(\ref{confmap1}) and~(\ref{confmap2}), the graphs of Figure~\ref{figeta3sde}
are reduced to $2$-point ones which are easier to evaluate.

\begin{figure}[ht]\centering
\includegraphics{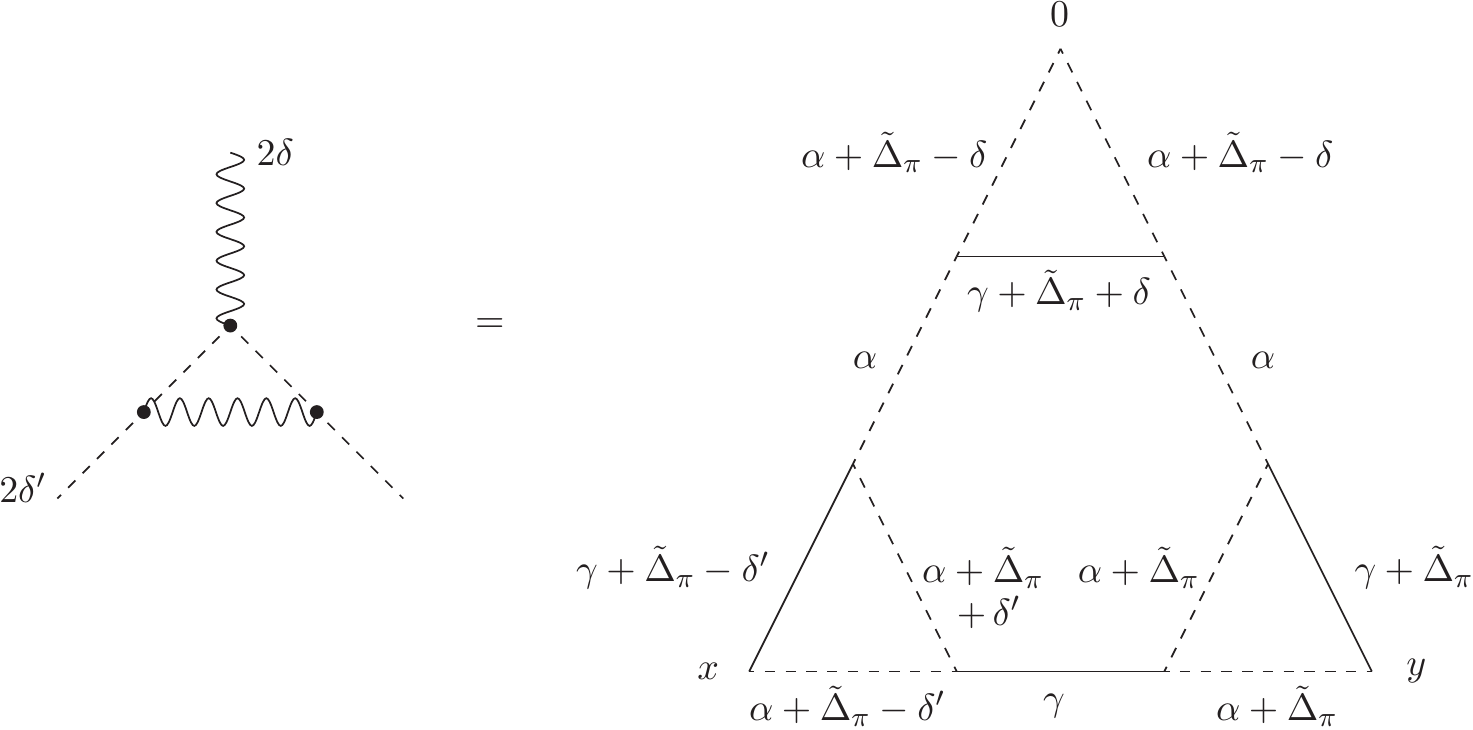}

\caption{Regularized one loop contribution to the vertex bootstrap equations.}\label{figreg}
\end{figure}

The graphs of Figure~\ref{figeta3sde}, however, are the lowest order
contributions to the full vertex function that we will denote by
$V(\bar{y},\alpha,\gamma;\delta,\delta^\prime)$, where the last two arguments
are regularizing parameters. These are required in the derivation of one of the
two consistency equations defining the large~$N$ bootstrap formalism
\cite{32,30,31,14}. The first equation represents Figure~\ref{figeta3sde} and
is
\begin{equation}
1 = V(\bar{y},\alpha,\gamma;0,0),
\label{cnfb1}
\end{equation}
where $\bar{y}$ is similar to the early amplitude combination $y$ but includes
the normalization of the Polyakov conformal triangle of Figure~\ref{figcntr}.
Indeed~(\ref{cnfb1}) is responsible for determining the terms in the $1/N$
expansion of $\bar{y}$ once the first few orders of $\eta$ have been reproduced
in this formalism. This is because the explicit expressions are necessary to
extract $\eta_3$ from the third order term of~the second bootstrap equation
which is
\begin{gather}
\frac{T_F N r(\alpha-1)}{C_F p(\gamma)} =
\frac{ \big[ 1 + 2 \chi_\pi \frac{\partial }{\partial \delta^\prime}
V(\bar{y},\alpha,\gamma;\delta,\delta^\prime) \big]}
{ \big[ 1 + 2 \chi_\pi \frac{\partial }{\partial \delta}
V(\bar{y},\alpha,\gamma;\delta,\delta^\prime) \big]}
\bigg|_{\delta=\delta^\prime=0} .
\label{cnfb2}
\end{gather}
We note briefly that the regularizations $\delta$ and $\delta^\prime$ that
appear here arise because of singularities in the $2$-point Schwinger--Dyson
equations when all the vertices are replaced by conformal triangles. In other
words it was recognised in the original work of~\cite{33} that in the absence
of any regularization the $2$-point functions with dressed vertices would be
finite overall. However each of the contributing diagrams were individually
divergent. To accommodate this, and similar to the introduction of $\Delta$
earlier, the vertex anomalous dimension has to be continued in a parallel way
to~(\ref{chiDel}). This is illustrated in Figure~\ref{figreg} for the leading
order contribution to $V(\bar{y},\alpha,\gamma;\delta,\delta^\prime)$, where we
have set
\begin{equation*}
\chi_\pi = 2 \tilde{\Delta}_\pi
\end{equation*}
for shorthand. This figure indicates the values of the exponents of the
internal lines of the Polyakov triangle. Moreover the appearance of both
$\delta$ and $\delta^\prime$ on the external legs of the graph on the left hand
side indicate the addition of the regularizations to the exponents of the
respective fields. This is reflected internally in the conformal triangles in
the right hand graph as each of~the vertices have to be unique even when there
is a regularization.

Given the form of~(\ref{cnfb2}) we can rederive $\eta_2$ from knowledge of the
value of the graph of~Figure~\ref{figreg}. This was determined in~\cite{26},
using the technique given in~\cite{14}, for the Ising Gross--Neveu model as a
function of the exponents of that theory and that result can be translated to~(\ref{lagchgnuniv}). If~we expand
$V(\bar{y},\alpha,\gamma;\delta,\delta^\prime)$ in large $N$ in the same
notation as~(\ref{expexp}) then to the order that will eventually be necessary
to evaluate $\eta_3$ we recall~\cite{26}
\begin{equation*}
V_1 = - \frac{Q^3}{\tilde{\Delta}_\pi \big(\tilde{\Delta}_\pi-\delta\big)
\big(\tilde{\Delta}_\pi-\delta^\prime\big)}
\exp\big [ F(\delta, \delta^\prime,\tilde{\Delta}_\pi)\big]
\end{equation*}
with
\begin{equation*}
Q = - \frac{a^2(\alpha-1) a(\gamma)}{(\alpha-1)^2\Gamma(\mu)}
\end{equation*}
and
\begin{gather}
F\big(\delta,\delta^\prime,\tilde{\Delta}_\pi\big) =\bigg[ 5 B_\gamma - 2 B_{\alpha-1} - 3 B_0
- \frac{2}{\alpha-1}\bigg]\tilde{\Delta}_\pi- [B_\gamma - B_0] \delta
\nonumber
\\ \hphantom{F\big(\delta,\delta^\prime,\tilde{\Delta}_\pi\big) =}
{}+ \bigg[ B_0 - B_{\alpha-1} - \frac{1}{\alpha-1} \bigg]\delta^\prime
+ \bigg[ C_{\alpha-1} - \frac{1}{\alpha-1} \bigg]\delta \delta^\prime
\nonumber
\\ \hphantom{F\big(\delta,\delta^\prime,\tilde{\Delta}_\pi\big) =}
{}+\bigg[ C_\gamma + C_0 - 2 C_{\alpha-1} + \frac{2}{(\alpha-1)^2}\bigg] \tilde{\Delta}_\pi\delta
{}+ \frac{1}{2} \bigg[ \frac{1}{(\alpha-1)^2} - C_{\alpha-1}- C_0 \bigg] {\delta^\prime}^2
\nonumber
\\ \hphantom{F\big(\delta,\delta^\prime,\tilde{\Delta}_\pi\big) =}
+ \bigg[ C_0 - C_\gamma - 2 C_{\alpha-1} + \frac{2}{(\alpha-1)^2}\bigg]
\tilde{\Delta}_\pi \delta^\prime- \frac{1}{2} [C_\gamma + C_0] \delta^2
\nonumber
\\ \hphantom{F(\delta,\delta^\prime,\tilde{\Delta}_\pi) =}
{}+ \bigg[ C_{\alpha-1} - \frac{7}{2} C_\gamma - \frac{3}{2} C_0
- \frac{1}{(\alpha-1)^2} \bigg]\tilde{\Delta}_\pi^2 +
O \big(\tilde{\Delta}_\pi^3, \delta^3, {\delta^\prime}^3 \big),
\label{cfv1}
\end{gather}
where the order symbol indicates that terms cubic in any combination of the
parameters are neglected. The functions $B_z$ and $C_z$ are defined by~\cite{14}
\begin{gather*}
B_z= \psi(\mu-z) + \psi(z),\qquad B_0 = \psi(1) + \psi(\mu),
\\
C_z = \psi^\prime(z) - \psi^\prime(\mu-z),\qquad C_0 = \psi^\prime(\mu) - \psi^\prime(1)
\end{gather*}
in terms of the Euler $\psi$ function. With~(\ref{cfv1}) and setting
$V=V_1/N$ in~(\ref{cnfb1}) and~(\ref{cnfb2}) it is straightforward to
expand both bootstrap equations to $O(1/N)$ and verify the earlier expressions
for $\eta_1$ and $\eta_2$.

Having established the formalism reproduces available results the extension to
the next order to find $\eta_3$ requires several steps. The first is to compute
the next term in the $1/N$ expansion of~(\ref{cfv1}) as that will contribute
to the $O\big(1/N^2\big)$ part of~(\ref{cnfb2}). However contributions from all the
graphs of Figure~\ref{figeta3sde} bar the one loop one have to be determined
and included as they correspond to $V_2$. The relevant parts of these graphs
were calculated in~\cite{26}. By this we mean that a~com\-pu\-ta\-tional shortcut
was used akin to that used in~\cite{14}. From examining the terms in the formal
Taylor expansion of~(\ref{cnfb2}) in powers of $1/N$ the part involving $V_2$
occurs in the combination
\begin{gather}
\bigg[\frac{\partial}{\partial \delta^\prime}V_2(y,\alpha,\gamma;\delta,\delta^\prime) -
\frac{\partial}{\partial \delta}V_2(y,\alpha,\gamma;\delta,\delta^\prime) \bigg]
\bigg|_{\delta=\delta^\prime=0} .
\label{v2diff}
\end{gather}
In~\cite{26} the contribution to~(\ref{v2diff}) from each of the higher order
correction graphs in Figure~\ref{figeta3sde} were determined. Indeed in most
cases only the value for the difference in derivatives could be found.
Appending the group theory values from the {\tt color.h} routine allows us to
finally extract $\eta_3$. We~find
\begin{gather*}
\eta_3 = \bigg[
\frac{(2 \mu-1) \big(35 \mu^3-43 \mu^2+16 \mu-2\big)}{4 \mu^2 (\mu-1)^4} C_F^2
- \frac{\mu^2 \big(43 \mu^2-35 \mu+6\big)}{8 \mu^2 (\mu-1)^4} C_F C_A
 \nonumber
 \\ \hphantom{\eta_3 =\bigg[}
{}- \frac{\mu^4 \big(4 \mu^3-5 \mu^2-9 \mu-14\big)}{48 \mu^2 (\mu-1)^4} C_A^2
- \frac{\mu^4 \big(4 \mu^3-2 \mu^2-3 \mu+10\big)}{4 \mu^2 (\mu-1)^4}
 \frac{d_F^{abcd} d_F^{abcd}}{T_F^2 N_A}
\nonumber
\\ \hphantom{\eta_3 =\bigg[}
{}+ \bigg[ \frac{1}{2} (11 \mu-3) (2 \mu-1)^2- \frac{1}{4} \mu (19 \mu-2) (2 \mu-1) C_F C_A
- \frac{1}{24} \mu^3 \big(2 \mu^2-6 \mu-23\big) C_A^2
\nonumber
\\ \hphantom{\eta_3 =\bigg[+\bigg[}
{}- \frac{ \mu^3 \big(2 \mu^2-3 \mu+4\big) d_F^{abcd} d_F^{abcd}}{2 T_F^2 N_A} \bigg]
\frac{\Psi(\mu)}{\mu (\mu-1)^3}
\\ \hphantom{\eta_3 =\bigg[}
{}+ \frac{3 ( 2 C_F - 4 \mu C_F + \mu C_A )^2}{8 (\mu-1)^2} \Psi^2(\mu)
+ \frac{( 2 C_F - 4 \mu C_F + \mu C_A )^2}{8 (\mu-1)^2} \Phi(\mu)
\nonumber
\\ \hphantom{\eta_3 =\bigg[}
{}+ \bigg[ (2 \mu-1) (\mu+1) C_F^2- \frac{1}{2} \mu (5 \mu-1) C_F C_A
- \frac{1}{24} \mu^2 (\mu-4) C_A^2
\\ \hphantom{\eta_3 =\bigg[+\bigg[}
{}+ \mu^2 (\mu+8) \frac{d_F^{abcd} d_F^{abcd}}{T_F^2 N_A} \bigg]
\bigg[\Theta(\mu) + \frac{1}{(\mu-1)^2}\bigg] \frac{1}{4(\mu-1)^2}
\nonumber
\\ \hphantom{\eta_3 =\bigg[}
{}-\bigg[ C_A^2 - 24 \frac{d_F^{abcd} d_F^{abcd}}{N_A T_F^2} \bigg]
\bigg[ \Theta(\mu) + \frac{1}{(\mu-1)^2} \bigg][ 2 \Psi(\mu) + \Xi(\mu) ]
\frac{\mu^2}{16 (\mu-1)}\bigg] \frac{\eta_1^3}{C_F^2} .
\end{gather*}
Again the pieces arising from the light-by-light graphs can be clearly
identified. In addition to the Euler $\psi$ function and its derivatives a new
function arises which is related to the function~$I(\mu)$ defined in~\cite{14}.
It corresponds to the derivative of a two loop self-energy graph, where a~regularizing exponent of one of the propagators is differentiated. In~\cite{35}
this master graph was expressed in terms of an ${}_4F_3$ hypergeometric
function where the regularizing parameter appears in the parameter arguments of
the function. A compact expression for $I(\mu)$ that is valid in~$d$-dimensions
was recorded in~\cite{36}. Here we have set
\begin{equation*}
I(\mu) = - \frac{2}{3(\mu-1)} + \Xi(\mu),
\end{equation*}
so that $\Xi(\mu)$ has an $\epsilon$ expansion involving multiple zeta
values~\cite{37,35,14}. For instance,
\begin{equation*}
\Xi(1-\epsilon) = \frac{2}{3} \zeta_3 \epsilon^2 + \zeta_4 \epsilon^3 +
\frac{13}{3} \zeta_5 \epsilon^4 + O\big(\epsilon^5\big) .
\end{equation*}
In strictly three dimensions~\cite{14}
\begin{equation*}
I\bigg(\frac32\bigg) = 2 \ln 2 + \frac{3\psi^{\prime\prime}\big(\half\big)}{2\pi^2}
\end{equation*}
and the expansion around three dimensions is known up to ten terms~\cite{38}.

\section{Results}\label{sec7}

We focus in this section on general aspects of the critical exponents of
(\ref{lagchgnuniv}) that we have determined. One of the main reasons for
considering a generalized universality class was the fact that known results
for specific models could be extracted as well as be of use where a different
Lie group underlies the physics problem. To assist with that we have collected
electronic expressions in an attached data file. One aspect of the results that
needs to be stated is that we have checked that the $d$-dimensional exponents
are in agreement with several models where the results were determined
directly. For instance, the original Ising Gross--Neveu model of~\cite{1}
corresponds to the abelian limit of the symmetry group by specifying
\begin{equation*}
C_F = 1 ,\qquad
T_F = 1 ,\qquad
d_F^{abcd} d_F^{abcd} = 1 ,\qquad
C_A = 0
\end{equation*}
while the Mott insulating phase~\cite{8,9} that corresponds to taking the
symmetry group to be ${\rm SU}(2)$ takes the values
\begin{equation*}
C_F = \frac{3}{4} ,\qquad
T_F = \frac{1}{2} ,\qquad
d_F^{abcd} d_F^{abcd} = \frac{5}{64} ,\qquad
C_A = 2 .
\end{equation*}
For the more recent application of~(\ref{lagchgnuniv}) to the fractionalized
Gross--Neveu model discussed in~\cite{11} the respective values are
\begin{equation*}
C_F = 2 ,\qquad
T_F = 2 ,\qquad
d_F^{abcd} d_F^{abcd} = \frac{20}{3} ,\qquad
C_A = 2 .
\end{equation*}
For each of these cases the $\epsilon$ expansion of the exponents near four
dimensions with $d=4-2\epsilon$ are in full agreement with known
three and four loop perturbative results~\cite{40,43,39,41,44,11,42,15}. In~the case of the Ising Gross--Neveu model exponents these also are in accord with
two dimensional perturbation theory~\cite{47,48,50,51,46,49,45}. Moreover
taking the limits for the three cases, all the large $N$ exponents agree with
previous work~\cite{22,20,21,25,52,44,19,23,24}.

One advantage of the arbitrary group approach in $d$-dimensions means that the
structure of the exponents can be studied in various representations. For
example if we restrict the fermions to be in the adjoint representation $A$
whence $C_F=C_A$ and $T_F=C_A$. In that case we find
\begin{gather*}
\eta_2^{\rm adj} =
\bigg[ \frac{13 \mu^2-12 \mu+2}{4\mu[\mu-1]^2}+ \frac{3 \mu-2}{2[\mu-1]} \Psi(\mu) \bigg]
\big( \eta_1^{\rm adj} \big)^2,
\\
\chi_{\pi \,1}^{\rm adj} =\frac{\mu}{2[\mu-1]} \eta_1^{\rm adj},
\\
\chi_{\pi \, 2}^{\rm adj} =
\bigg[\bigg[\frac{3 \mu^2}{\mu-1} \frac{d_A^{abcd} d_A^{abcd}}{C_A^4 N_A}
- \frac{\mu^2}{8[\mu-1]} \bigg] \Theta(\mu)
+ \frac{\mu (3 \mu-2)}{4[\mu-1]^2} \Psi(\mu)
\\ \hphantom{\chi_{\pi \, 2}^{\rm adj} =\bigg[\bigg[}
{}- \frac{\mu (\mu+1) (\mu-3)}{6[\mu-1]^2}
- \frac{\mu^2 (2 \mu-1)}{[\mu-1]^2} \frac{d_A^{abcd} d_A^{abcd}}{C_A^4 N_A}
\bigg]\big( \eta_1^{\rm adj} \big)^2,
%\nonumber
\\
\lambda_2^{\rm adj} =\bigg[ \bigg[ \frac{1}{6} \mu \big(3 \mu^2-6 \mu+2\big)
+ 4 \mu \frac{d_A^{abcd} d_A^{abcd}}{C_A^4 N_A} \bigg]
\frac{1}{[\mu-1][\mu-2]^2 \eta_1^{\rm adj}}
\\ \hphantom{\lambda_2^{\rm adj} =\bigg[}
{}- \bigg[ \frac{1}{24} \mu^2 (2 \mu-3)
+ 2 \mu^2 (2 \mu-3) \frac{d_A^{abcd} d_A^{abcd}}{C_A^4 N_A} \bigg]
\frac{\Psi^2(\mu)+\Phi(\mu)}{[\mu-1][\mu-2]}
\\ \hphantom{\lambda_2^{\rm adj} =\bigg[}
{}- \bigg[ \frac{1}{24} (\mu-1) \big(96 \mu^5-438 \mu^4+549 \mu^3+4 \mu^2-288 \mu+96\big)
\\ \hphantom{\lambda_2^{\rm adj} =\bigg[{-}\bigg[}
{}+ \mu^2 (3 \mu-5) (2 \mu-5) \frac{d_A^{abcd} d_A^{abcd}}{C_A^4 N_A} \bigg]
\frac{\Psi(\mu)}{[\mu-1]^2[\mu-2]^2}
\\ \hphantom{\lambda_2^{\rm adj} =\bigg[}
{}+ \bigg[ 3 \mu^2 (5 \mu-7) \frac{d_A^{abcd} d_A^{abcd}}{C_A^4 N_A}
- \frac{1}{8} \mu^2 (\mu-2) (12 \mu-7) \bigg]
\frac{\Theta(\mu)}{[\mu-1][\mu-2]}
\\ \hphantom{\lambda_2^{\rm adj} =\bigg[}
{}+ \frac{192 \mu^8-1544 \mu^7+4770 \mu^6-6865 \mu^5+3951 \mu^4+724 \mu^3
-1896 \mu^2 +768 \mu-96}{48\mu[\mu-1]^3[\mu-2]^2}
\\ \hphantom{\lambda_2^{\rm adj} =\bigg[}
{}- \frac{\mu^2 \big(4 \mu^4-18 \mu^3+26 \mu^2-15 \mu+7\big)}{2[\mu-1]^3[\mu-2]^2}
\frac{d_A^{abcd} d_A^{abcd}}{C_A^4 N_A} \bigg]
\big( \eta_1^{\rm adj} \big)^2,
%\nonumber
\\
\eta_3^{\rm adj} =
\bigg[\frac{(3 \mu-2)^2}{8[\mu-1]^2} [ \Phi(\mu) + 3 \Psi^2(\mu) ]
- \frac{\mu^2 (2 \mu-1) (\mu^2-1)}{2[\mu-1]^4}\frac{d_A^{abcd} d_A^{abcd}}{C_A^4 N_A}
\\ \hphantom{\eta_3^{\rm adj} =\bigg[}
{}- \frac{8 \mu^7-10 \mu^6-17 \mu^5-1184 \mu^4+2448 \mu^3-1704 \mu^2+480 \mu-48}
{96\mu^2[\mu-1]^4}
\\ \hphantom{\eta_3^{\rm adj} =\bigg[}
{}- \bigg[ \frac{1}{12} \big(\mu^5-3 \mu^4-160 \mu^3+267 \mu^2-132 \mu+18\big)
\\ \hphantom{\eta_3^{\rm adj} =\bigg[{-} \bigg[ }
{}+ \frac{1}{2} \mu^3 (2 \mu+1) (\mu-2) \frac{d_A^{abcd} d_A^{abcd}}{C_A^4 N_A}
\bigg] \frac{\Psi(\mu)}{\mu[\mu-1]^3}
\\ \hphantom{\eta_3^{\rm adj} =\bigg[}
{}+ \bigg[\frac{1}{4} \mu^2 (\mu+8) \frac{d_A^{abcd} d_A^{abcd}}{C_A^4 N_A}
- \frac{1}{96} (\mu^3+8 \mu^2-36 \mu+24) \bigg] \frac{\Theta(\mu)}{[\mu-1]^2}
\\ \hphantom{\eta_3^{\rm adj} =\bigg[}
{}- \frac{\mu^2}{8[\mu-1]}
\bigg[ 1 - 24 \frac{d_A^{abcd} d_A^{abcd}}{C_A^4 N_A} \bigg]
\Theta(\mu) \Psi(\mu)- \frac{\mu^2 \Xi(\mu)}{16[\mu-1]^3}
\bigg[ 1 - 24 \frac{d_A^{abcd} d_A^{abcd}}{C_A^4 N_A} \bigg]
\\ \hphantom{\eta_3^{\rm adj} =\bigg[}
{}- \frac{\mu^2}{16[\mu-1]} \bigg[ 1 - 24 \frac{d_A^{abcd} d_A^{abcd}}{C_A^4 N_A}\bigg]
\Xi(\mu) \Theta(\mu)\bigg] \big( \eta_1^{\rm adj} \big)^3,
\end{gather*}
where $d_A^{abcd}$ is the adjoint version of the fully symmetric rank $4$
Casimir. These exponents simplify substantially in three dimensions since
\begin{gather*}
\lambda \big|_{d=3} = 1 - \frac{16}{3\pi^2 N} +
\bigg[ 96 \frac{d_A^{abcd} d_A^{abcd}}{C_A^4 N_A}+ \frac{5248}{\pi^2} - 432 \bigg]
\frac{1}{27\pi^2 N^2},
\\
\eta \big|_{d=3} = \frac{8}{3\pi^2 N} +
\frac{1216}{27\pi^4 N^2} + \big[ \big[ 9072 \zeta_3 - 864 \pi^2 \ln 2 \big]
\big[ C_A^4 N_A - 24 d_A^{abcd} d_A^{abcd} \big]
\\ \hphantom{\eta \big|_{d=3} =}
{}+ \big[ 25920 \pi^2 - 435456 \big] d_A^{abcd} d_A^{abcd}
+ \big[ 151072 - 8760 \pi^2 \big] C_A^4 N_A
\big] \frac{1}{243 \pi^6 C_A^4 N_A N^3}.
\end{gather*}
The group valued coefficient of the terms involving $\zeta_3$ and
$\pi^2 \ln 2$, which derive from $I\big(\threehalves\big)$, has an interesting
combination of Casimirs. Indeed there might be instances of this factor being
zero for certain Lie groups. However, we have computed the value of
$\big(C_A^4 N_A-24 d_A^{abcd} d_A^{abcd}\big)$ for all the classical and
exceptional Lie groups and found that it is always non-zero.

\section{Discussion}\label{sec8}

\looseness=1 As the Ising Gross--Neveu universality class is central to a number of phase
transitions in various materials, we have examined a generalized version of the
underlying quantum field theory that incorporates the respective condensed
matter systems. The key aspect is that the core interaction is endowed with a
non-abelian symmetry that has a parallel in gauge theories. There the gauge
interaction of QED is extended from an abelian to a non-abelian one to produce
QCD by the inclusion of the generators of a Lie group thereby endowing QED with
a colour symmetry. The similar extension of the Ising Gross--Neveu model is
simpler in some respects. One obvious one is the absence of gauge symmetry. A
benefit, however, is that considering~(\ref{lagchgnuniv}) at the outset means
results for specific phase transitions can be quickly deduced by specifying the
Lie group. Indeed if a phase transition were discovered in a material that was
in the same universality class as the Ising Gross--Neveu model but possessed a
new symmetry other than the specific examples we have noted here, then
information on the exponents can readily be deduced from our results.
Throughout we have focussed on the application of the critical point large $N$
technique developed in~\cite{13,12,14} to determine $d$-dimensional critical
exponents. The advantage of this is that results are available for the
renormalization group functions of the four dimensional quantum field theories
in the same universality class too. By the same token the large $N$ exponents
contain a wealth of information on the structure of the same functions. For~instance, coefficients in the anomalous dimension beyond the first few known
loop orders can be accessed at successive orders in $1/N$. This is particularly
useful in that our $O\big(1/N^2\big)$ and $O\big(1/N^3\big)$ exponents can reveal where the new
colour group Casimirs, such as $d_F^{abcd} d_F^{abcd}$, appear. In indicating
the parallel of the QED to QCD generalization, examining the large $N$
$O\big(1/N^2\big)$ exponents in this universality class, albeit with a simpler vertex
structure, does provide useful insight into what to expect in the calculation
of critical exponents in QCD at $O\big(1/N^2\big)$. We~have to qualify this comment by
noting that while there are similarities, in the QCD large $N$ critical
exponent computation for $\nu$ for instance, there will be more graphs to
consider than those of Figure~\ref{figlam2}. This is because in
(\ref{lagchgnuniv}) Feynman diagrams with subgraphs involving three $\pi^a$
lines connecting to a fermion loop are zero after taking the $\gamma$-matrix
trace. In QCD this would not be the case due to each vertex adding an extra
$\gamma$-matrix to the trace. While such graphs remain to be computed the
associated group theory factor that would result should not involve any Casimir
higher than $d_F^{abcd} d_F^{abcd}$.

\subsection*{Acknowledgements} This work was fully supported by a DFG Mercator
Fellowship and in part with the STFC Conso\-li\-dated ST/T000988/1. The graphs were
drawn with the {\sc Axodraw} package~\cite{53}. Computations were carried out
in part using the symbolic manipulation language {\sc Form}~\cite{28, 27}.

\pdfbookmark[1]{References}{ref}
\LastPageEnding

\end{document}